\newif\ifreport
\reporttrue

\documentclass[10pt,conference]{IEEEtran}
\IEEEoverridecommandlockouts

\usepackage{cite}
\usepackage{amsmath,amssymb,amsfonts}
\usepackage{graphicx}
\usepackage{textcomp}
\usepackage[dvipsnames]{xcolor}

\usepackage{paralist}\usepackage[numbers]{natbib}
\usepackage{ifthen}\usepackage{algorithm}
\usepackage{algorithmicx}
\usepackage[noend]{algpseudocode}
    \algrenewcommand\alglinenumber[1]{\scriptsize #1:}\usepackage{tikz}
\usetikzlibrary{calc}
\usepackage{xspace}

\usepackage{hyperref}\hypersetup{colorlinks=true,linkcolor=black,citecolor=black,urlcolor=black}
\usepackage{cleveref}
\crefname{subsection}{\S}{\S\S}
\crefname{subsubsection}{\S}{\S\S}
\crefrangeformat{line}{lines~#3#1#4-#5#2#6}
\Crefrangeformat{line}{Lines~#3#1#4-#5#2#6}
 \def\BibTeX{{\rm B\kern-.05em{\sc i\kern-.025em b}\kern-.08em
    T\kern-.1667em\lower.7ex\hbox{E}\kern-.125emX}}
 \newcommand{\QAICCC}{QAICCC\xspace}
\newcommand{\QAICCClong}{Qubit Allocation for Inter-Circuit Crosstalk Countermeasure\xspace}

\newcommand{\Python}{Python\xspace}
\newcommand{\Qiskit}{Qiskit\xspace}
\newcommand{\PennyLane}{PennyLane\xspace}
\newcommand{\pyGSTi}{pyGSTi\xspace}
\newcommand{\XtalkSched}{XtalkSched\xspace}
\newcommand{\ColorDynamic}{ColorDynamic\xspace}
 \newcommand{\quotes}[1]{\text{``{#1}''}}

\newcommand{\todoColorDefault}{Black}\newcommand{\todoColor}{\todoColorDefault}\newcommand{\todoSetColor}[1]{\ifthenelse{\equal{#1}{DB}}{
        \renewcommand{\todoColor}{Red}}{}\ifthenelse{\equal{#1}{YM}}{
        \renewcommand{\todoColor}{BurntOrange}}{}}

\DeclareMathOperator{\cardSymb}{card}
\newcommand{\card}[1]{\cardSymb({#1})}
\newcommand{\knowing}{\,|\,}
\newcommand{\set}[2]{\left\{{#1}\ifthenelse{\equal{\unexpanded{#2}}{}}{}{\knowing {#2}}\right\}}
\newcommand{\dataList}[1]{\left({#1}\right)}

\algnewcommand\algorithmicinput{\textbf{input:}}
\algnewcommand\AlgoInput{\item[\algorithmicinput]}
\algnewcommand\algorithmicoutput{\textbf{output:}}
\algnewcommand\AlgoOutput{\item[\algorithmicoutput]}
\algnewcommand\algorithmictype{\textbf{type:}}
\algnewcommand\AlgoType{\item[\algorithmictype]}
\DeclareMathOperator{\algoError}{Error}
\DeclareMathOperator{\algoTrue}{True}

\newcommand{\xTalk}[2]{${#1}$-${#2}$}
\newcommand{\xTalkImpacting}{Q_\rightarrow}
\newcommand{\xTalkImpacted}{Q_\leftarrow}
\newcommand{\xTalkScore}{\mathit{score}}
\newcommand{\xTalkScores}{\mathit{ErrorRates}}
\newcommand{\xTalkTriple}{\mathit{rate}}

\newcommand{\typeAlloc}{\text{Allocation}\xspace}
\newcommand{\typeAllocs}{\text{Allocations}\xspace}

\newcommand{\Sizes}{\mathit{Sizes}}
\newcommand{\Trusted}{Q_T}
\newcommand{\Untrusted}{Q_U}
\newcommand{\Unallocated}{Q_\varnothing}
\newcommand{\Population}{\mathit{Pop}}
\newcommand{\CurrentPop}{\mathit{CurrentPop}}
\newcommand{\Candidates}{\mathit{Archive}}
\newcommand{\allocation}{\mathit{alloc}}
\newcommand{\Allocations}{\mathit{Allocs}}
\newcommand{\allocScoreName}{score}
\newcommand{\allocPenaltyName}{penalty}
\newcommand{\allocIncidentalName}{Incidental}
\newcommand{\xTalkLastName}{lastRate}
\DeclareMathOperator{\allocScoreSymb}{\allocScoreName}
\DeclareMathOperator{\allocPenaltySymb}{\allocPenaltyName}
\DeclareMathOperator{\allocIncidentalSymb}{\allocIncidentalName}
\DeclareMathOperator{\xTalkLastSymb}{\xTalkLastName}

\newcommand{\allocIncidental}[1]{\allocIncidentalSymb({#1})}
\newcommand{\xTalkLast}[1]{\xTalkLastSymb({#1})}
\DeclareMathOperator{\isFixedSymb}{isSafe}
\newcommand{\isFixed}[2]{\isFixedSymb({#1}, {#2})}
\newcommand{\evalAllocName}{updAlloc}
\DeclareMathOperator{\evalAllocSymb}{\evalAllocName}
\newcommand{\evalAlloc}[2]{\evalAllocSymb({#1}, {#2})}

\DeclareMathOperator{\nbUsersSymb}{nbUsers}
\newcommand{\nbUsers}[2]{\nbUsersSymb({#1}, {#2})}
\newcommand{\updSizesName}{updSizes}
\DeclareMathOperator{\updSizesSymb}{\updSizesName}
\newcommand{\updSizes}[1]{\updSizesSymb({#1})}
\newcommand{\SizeTrusted}{S_T}
\newcommand{\SizeUntrusted}{S_U}
\newcommand{\initAllocName}{initAlloc}
\DeclareMathOperator{\initAllocSymb}{\initAllocName}
\newcommand{\initAlloc}[1]{\initAllocSymb({#1})}
\newcommand{\improveAllocName}{allocMerge}
\DeclareMathOperator{\improveAllocSymb}{\improveAllocName}
\newcommand{\improveAlloc}[1]{\improveAllocSymb({#1})}
\newcommand{\archiveAllocName}{archAlloc}
\DeclareMathOperator{\archiveAllocSymb}{\archiveAllocName}
\newcommand{\archiveAlloc}[1]{\archiveAllocSymb({#1})}
\newcommand{\updPopName}{updPop}
\DeclareMathOperator{\updPopSymb}{\updPopName}
\newcommand{\updPop}[1]{\updPopSymb({#1})}

\newcommand*\getscale[1]{\begingroup
    \pgfgettransformentries{\scaleA}{\scaleB}{\scaleC}{\scaleD}{\whatevs}{\whatevs}\pgfmathsetmacro{#1}{sqrt(abs(\scaleA*\scaleD-\scaleB*\scaleC))}\expandafter
  \endgroup
  \expandafter\def\expandafter#1\expandafter{#1}}
\makeatletter
\newdimen\@XCoord
\newdimen\@YCoord
\newdimen\XCoordA
\newdimen\YCoordA
\newcommand*{\ExtractCoordinateA}[1]{\getscale{\@scalefactor}\path [transform canvas] (#1); \pgfgetlastxy{\@XCoord}{\@YCoord};\pgfmathsetlength{\XCoordA}{\@XCoord/\@scalefactor}\pgfmathsetlength{\YCoordA}{\@YCoord/\@scalefactor}}
\makeatother
\makeatletter
\newdimen\@XCoord
\newdimen\@YCoord
\newdimen\XCoordB
\newdimen\YCoordB
\newcommand*{\ExtractCoordinateB}[1]{\getscale{\@scalefactor}\path [transform canvas] (#1); \pgfgetlastxy{\@XCoord}{\@YCoord};\pgfmathsetlength{\XCoordB}{\@XCoord/\@scalefactor}\pgfmathsetlength{\YCoordB}{\@YCoord/\@scalefactor}}
\makeatother

\newcommand{\offsetBox}{0.08}
\newcommand{\offsetQubit}{0.25}
\newcommand{\xBetween}{0.4}
\newcommand{\xSpace}{1.75}
\newcommand{\ySpace}{1.25}

\newcommand{\allocTrustedName}{allocTrusted}
\DeclareMathOperator{\allocTrustedSymb}{\allocTrustedName}
\newcommand{\allocTrusted}[1]{\allocTrustedSymb({#1})}
\newcommand{\allocUnallocatedName}{allocImpacted}
\DeclareMathOperator{\allocUnallocatedSymb}{\allocUnallocatedName}
\newcommand{\allocUnallocated}[1]{\allocUnallocatedSymb({#1})}
\newcommand{\allocImpactedName}{allocControl}
\DeclareMathOperator{\allocImpactedSymb}{\allocImpactedName}
\newcommand{\allocImpacted}[1]{\allocImpactedSymb({#1})}
\newcommand{\connectName}{connect}
\DeclareMathOperator{\connectSymb}{\connectName}
\newcommand{\connect}[1]{\connectSymb({#1})}
\newcommand{\User}{\mathit{User}}
\newcommand{\UserMerge}{\User_\mathit{merge}}
\newcommand{\maxLength}{\mathit{maxLen}}
\newcommand{\userPath}{\mathit{Path}}
\DeclareMathOperator{\remain}{rem}
\DeclareMathOperator{\Paths}{getPaths}
\newcommand{\newAllocName}{newAlloc}
\DeclareMathOperator{\newAlloc}{\newAllocName}
\DeclareMathOperator{\getUser}{getUser}
 
\begin{document}

\ifreport
\title{A Piece of QAICCC: Towards a Countermeasure Against Crosstalk
  Attacks in Quantum Servers}
\else
\title{A Piece of QAICCC: Towards a Countermeasure Against Crosstalk
  Attacks in Quantum Servers \\{\large NIER paper - QSYS track}
}\fi

\author{\IEEEauthorblockN{Yoann~Marquer}
\IEEEauthorblockA{\textit{University of Luxembourg}\\
Luxembourg \\
yoann.marquer@uni.lu, 0000-0002-4607-967X}
\and
\IEEEauthorblockN{Domenico~Bianculli}
\IEEEauthorblockA{\textit{University of Luxembourg}\\
Luxembourg \\
domenico.bianculli@uni.lu, 0000-0002-4854-685X}
}

\maketitle

\begin{abstract}
Quantum computing, while allowing for processing information exponentially faster than classical computing, requires computations to be delegated to quantum servers, which makes security threats possible.
For instance, previous studies demonstrated that crosstalk between
attacker and victim's qubits can be exploited to mount security
attacks.

In this idea paper, we propose the QAICCC approach to allocate qubits between users to minimize inter-circuit crosstalk and, thus, possibilities for attacks, while maximizing qubit usage.
Also, combined with existing techniques, QAICCC aims to reduce intra-circuit noise.
Thus, QAICCC will support quantum computing adoption by securing the usage of quantum servers by a large number of actors.
\end{abstract}


 \section{Introduction}
\label{sec:introduction}

Quantum computing (QC) uses quantum bits (qubits) instead of the bits used in classical computing, enabling massive parallel computation based on quantum physics properties like quantum superposition.
This allows quantum computers to process information exponentially faster than any classical computer, with empirical evidence for quantum supremacy~\cite{AAB+19}.
Thus, QC has an impact on many emerging technologies~\cite{AKM+24} and
industrial use cases, especially regarding optimization problems~\cite{LTH+24}.
For this reason, technology giants such as IBM, Google, and Microsoft have heavily invested in
QC~\cite{KAL24}.

Nevertheless, because of cost and needs for ultra-cold tempe\-rature, shielded environment, and complex wiring for control, QC is far from becoming a personal commodity~\cite{SAP+21}.
Hence, access to quantum computers is likely to be provided remotely, through \emph{quantum servers}.
Sharing quantum hardware between multiple users allows to efficiently use quantum resources, but make some security threats possible~\cite{GUS23}.
For instance, correlations between attacker and victim's qubits used in the same server can be exploited in various security attacks~\cite{SAP+21} like fault-injection attacks to disrupt victim's output~\cite{ASA+20} and data-leakage attacks to retrieve victim's output~\cite{ASG21}.

Several countermeasures have been designed against the unintended interaction of qubits, called \emph{crosstalk}~\cite{MMM+20}. 
Buffer qubits between user circuits could likely prevent crosstalk attacks~\cite{ASA+20},
but would be a waste of quantum resources.
Other approaches based on gate scheduling~\cite{MMM+20} or graph
coloring~\cite{DGL+20} have been used to reduce crosstalk in quantum circuits, but never between user circuits.

In this idea paper, we address the problem of securing user circuits
submitted to quantum servers, so that quantum computing can be used in a safe way by more and more actors.
We propose the \emph{\QAICCClong (\QAICCC)} approach, which
aims to maximize qubit usage, reduce inter-circuit crosstalk as
well as intra-circuit noise.
\QAICCC  performs (1) a crosstalk analysis of the targeted platform\footnote{In this
article, we use the term \quotes{platform} to indicate the chosen quantum
processing unit performing the computation (e.g.,
the \texttt{ibmqx2} platform represented in
\Cref{fig:qubitConnectivity} and IBM available
platforms~\cite{ibmQPUs}) or a simulator.}
to determine qubits involved in crosstalk with the largest intensity;
(2) the allocation of qubits in the safest possible way; 
and (3) the application of existing techniques to further reduce crosstalk between circuits and noise in user circuits.
The main contribution of this paper is the allocation algorithm, which
aims to maximize qubit usage (including making unused qubits available
for future usage) while minimizing the largest inter-circuit crosstalk error rate.

 \section{Background}
\label{sec:background}

\subsection{Crosstalk}

\begin{figure}[t]
\centering
\begin{tikzpicture}[scale=1.09]
\tikzset{
        vertex/.style={draw,circle},
        edge/.style={thick,color=black}
    }
\node[vertex] (0)
    	at (-1,0)
	{$q_0$};
    \node[vertex] (1)
    	at (0,1)
	{$q_1$};
    \node[vertex] (2)
    	at (0,0)
	{$q_2$};
    \node[vertex] (3)
    	at (1,0)
	{$q_3$};
    \node[vertex] (4)
    	at (0,-1)
	{$q_4$};
\draw[edge]
    	(0) -- (1);
    \draw[edge]
    	(0) -- (2);
    \draw[edge]
    	(1) -- (2);
    \draw[edge]
    	(2) -- (3);
    \draw[edge]
    	(2) -- (4);
    \draw[edge]
    	(3) -- (4);
\end{tikzpicture} \caption{Qubit connectivity of the \texttt{ibmqx2} platform.}
\label{fig:qubitConnectivity}
\end{figure}
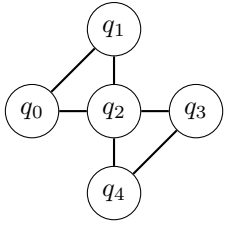

The qubit \emph{connectivity} of a quantum platform determines how
qubits are connected to each other. It is represented using
a graph, where qubits are vertices and edges indicates which qubits
are connected and thus can be used together by quantum gates to
perform quantum operations.
\Cref{fig:qubitConnectivity} illustrates the qubit
connectivity of an IBM 5-qubit platform; we will use this as running example.

Before execution, quantum programs must be \emph{transpiled}
1) to add swap gates to match platform connectivity and
2) to simulate unsupported gates by combining supported ones~\cite{PP24}.
Transpilation is usually done automatically by frameworks like
\Qiskit.

Qubits suffer from a short lifetime, leading to a spontaneous loss of qubit state information, called \emph{decoherence}, because of relaxation or dephasing~\cite{DGL+20}.
Another source of noise in quantum computing is the unintended interaction of qubits, called \emph{crosstalk}, which covers a wide range of physical phenomena and varies across quantum platforms~\cite{SPR+20}.
Crosstalk is usually due to qubit connectivity (adjacent qubits, like
$q_2$ and $q_3$ in \Cref{fig:qubitConnectivity},  are more likely to interact) and operating frequency (qubits tuned to close frequencies can be in resonance~\cite{DGL+20}).
Noise (including crosstalk) is measured using various metrics called \emph{error rates} (e.g., Hamiltonian, Stochastic, and Affine error rates) that can be obtained using tools like the \pyGSTi simulator~\cite{pyGSTi}.
For instance, the crosstalk error rate of  qubit $q_2$ in
\Cref{fig:qubitConnectivity} is larger when qubits $q_3$ and $q_4$ are
activated, because $q_2$ is connected to $q_3$ and $q_4$, and because $q_2$'s operating frequency is close to $q_4$'s one~\cite{ASA+20}.

\citet{ASA+20} demonstrated that crosstalk can be a larger source of
noise than quantum gate error and decoherence.
Moreover, they demonstrated that crosstalk between qubits allocated to different users can be exploited for security attacks.
In their attack scenario, a user (the victim) requests qubits to the server; then, another user (the attacker) requests several copies of small quantum circuits to control the largest number possible of remaining qubits.
The victim runs her circuit to obtain a result in a qubit output, e.g., $q_2$ in \Cref{fig:qubitConnectivity};
the attacker runs her circuit, e.g., a sequence of CNOT gates involving $q_3$ and $q_4$.
Due to the presence of crosstalk from $q_3$ and $q_4$ to $q_2$, the
magnitude of the victim's expected output can be reduced below the
magnitude of other results.
In other words, because of crosstalk, an attacker can use qubits (like
$q_3$ and $q_4$ in the example)
impacting a victim's qubit ( $q_2$ in the example) to tamper its quantum state and even the victim's outcome, after the impacted qubit is measured.
We call \emph{crosstalk attacks} security attacks made possible by
crosstalk, like \citet{ASA+20}'s fault injection attack.

\subsection{Noise reduction techniques}
Current quantum computers are called NISQ (noisy intermediate-scale quantum) computers, because they are larger than small-scale prototypes with a few qubits, but not large enough so that quantum error correction can be applied~\cite{LTH+24}.
Hence, current quantum executions are noisy, which decreases their \emph{success rate}, i.e., the probability to obtain the expected output~\cite{DGL+20}.

Many approaches have been proposed to reduce noise in quantum
circuits; below we briefly present those used in our approach.
Notably, \citet{MMM+20} proposed an approach to reduce the time required to identify crosstalk by considering only adjacent pairs (as crosstalk is usually a short-range phenomenon) and performing the measurements of distant pairs in parallel.
Moreover, they proposed a gate scheduling technique to reduce both decoherence and crosstalk, obtaining the \XtalkSched scheduler, which writes slightly longer circuits than the current state of the art in {\Qiskit} but outperformed it in terms of error rate.
\citet{DGL+20} proposed another approach to mitigate crosstalk, called \ColorDynamic, for allocating qubit and gate frequencies.
They considered the vertex-coloring of the connectivity graph of the targeted platform and the edge-coloring of a crosstalk graph representing relevant qubits pairs, in order to prevent close qubits or gates to share close operating frequencies and thus be in resonance.

 \section{Approach}
\label{sec:approach}

\subsection{Motivations and goals}
\label{sec:goals}

As NISQ computers do not have enough qubits for error correction, they resort to noisy computations that hinder QC adoption.
This situation warrants 1) increasing the number of available qubits
and 2) reducing noise in quantum computations.

For the former, since each qubit is a precious resource that should not be wasted,
we see \emph{maximizing qubit usage}---as in allocating qubits in quantum
servers to users or, if current users's needs are already satisfied, being ready for the next user(s)---as a priority.

For the latter, since crosstalk attacks are possible between users of the same quantum servers, in this paper we distinguish two kinds of noise:
\emph{inter-circuit crosstalk}, i.e., crosstalk between qubits allocated to different users, and \emph{intra-circuit noise}, i.e., crosstalk and decoherence involving qubits allocated to the same user.

Another priority for QC adoption is that quantum servers can be used in a safe way, without a user being able to interfere with other users' executions.
Hence the need to \emph{reduce inter-circuit crosstalk}.
More precisely, since crosstalk is quantified using error rate and the targeted platform may exhibit many qubit combinations leading to crosstalk, we aim to minimize the largest inter-circuit crosstalk error rate.

Finally, maximizing qubit usage while reducing inter-circuit crosstalk
implies that each combination of qubits involved in crosstalk should
be controlled, when possible, by the same user. Such a qubit
allocation would tend to increase intra-circuit noise, which should be
reduced as well to improve success rate and thus quality of service.
Therefore, \emph{intra-circuit noise reduction} is another priority in NISQ computers.

To summarize, our approach aims to achieve the following goals, in decreasing priority:
\begin{enumerate}[\bfseries G1:]
\item maximizing qubit usage (including making unused qubits available for future usage);
\item minimizing the largest inter-circuit crosstalk error rate;
\item reducing intra-circuit noise (crosstalk and decoherence).
\end{enumerate}

\subsection{Overview}

\begin{figure}[t]
\centering
\includegraphics[width=1.0\linewidth]{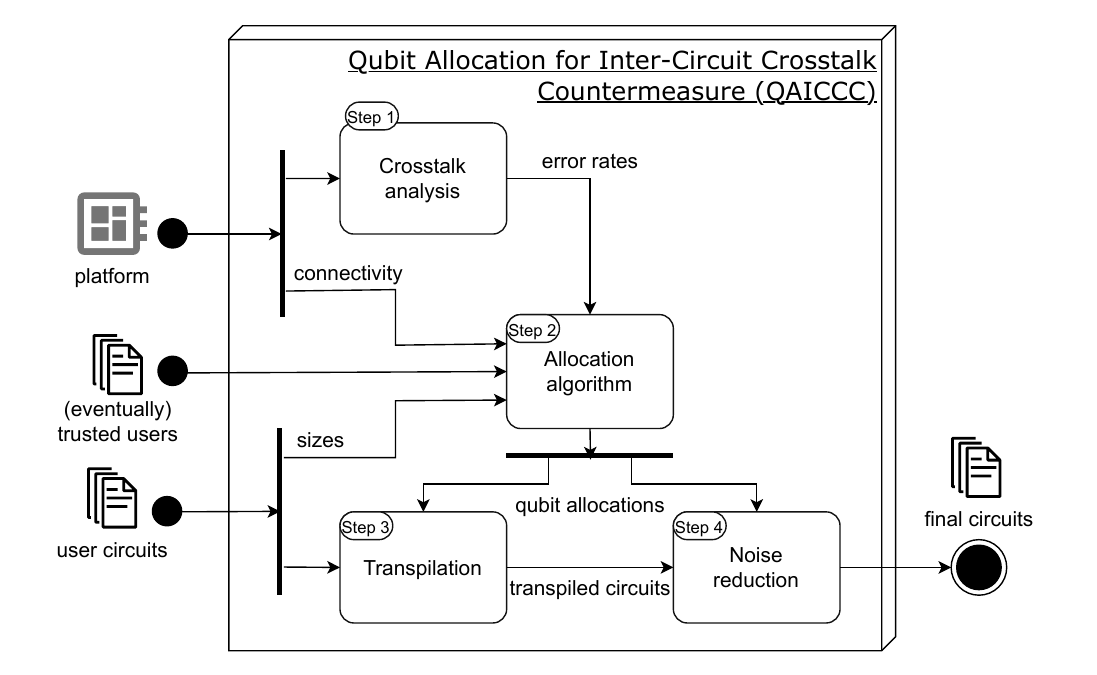}
\caption{Activity diagram of the \QAICCClong (\QAICCC) approach.}
\label{fig:approachDiagram}
\end{figure}

\Cref{fig:approachDiagram} provides a graphical overview of our
\QAICCClong (\QAICCC) approach, using the UML activity diagrams notation.
It takes as input the targeted quantum platform, the user circuits, and a
(potentially empty) list of trusted users; it returns transpiled
circuits (i.e., ready to be executed on the platform) that can be
executed in a safe way, minimizing the threat of crosstalk attacks.

The approach consists of four main steps.
In Step~1, the platform is
analyzed to determine crosstalk error rates (\Cref{sec:crosstalk}).
In Step~2, based on these error rates, the connectivity of the
platform, the sizes of the user circuits, and the list of trusted users, qubits are allocated to users;
we detail the allocation algorithm in \Cref{sec:allocation}.
In Step~3, user circuits are transpiled according to the selected qubit allocation to match the platform connectivity and supported gates (\Cref{sec:transpilation}).
In Step~4, transpiled circuits are transformed to reduce noise during
quantum executions, using existing techniques like \XtalkSched and \ColorDynamic
(\Cref{sec:noiseReduction}).

The \QAICCC approach fulfills the goals identified in \Cref{sec:goals} as follows.
To meet G1, if there are unused qubits, these are allocated so that
they are connected to each other, making them available to future users.
This is done in Step~2 by introducing a new (idle) user, allocating
the unallocated qubits  to the idle user, and ensuring that each user's qubit allocation forms a connected component.
To meet G2, qubits are allocated to minimize the largest inter-circuit crosstalk error rate.
This is done through Step~1, by quantifying error rates, Step~2, by processing them in decreasing order, and through Step~4, by applying  \XtalkSched and \ColorDynamic to reduce remaining inter-circuit crosstalk.
To meet G3,  \XtalkSched and \ColorDynamic are applied in Step~4 to minimize intra-circuit noise, while preserving the changes required to meet G2.

\subsection{Step 1: Crosstalk analysis}
\label{sec:crosstalk}

In this step, the platform is analyzed (e.g., with the \pyGSTi simulator~\cite{pyGSTi}) to determine crosstalk error rates between qubits.
We consider the impact of one or two qubits on another qubit (as in \citet{ASA+20}'s work) and the impact of two qubits on two qubits (i.e., gate errors, as in \citet{MMM+20}'s work), which we respectively call \xTalk{1}{1}, \xTalk{2}{1}, and \xTalk{2}{2} crosstalk.

To save time, we take inspiration from \citet{MMM+20}'s methodology (\Cref{sec:background}) by considering neighboring qubits and measuring error rates of distant qubits in parallel.
More precisely, each \xTalk{m}{n} crosstalk is measured so that the $m + n$ considered qubits form a connected component.
Note that this implies the existence of a path between each two
considered qubits; at the same time, it does not imply that all the
considered qubits are directly connected to each other.
For instance, in \Cref{fig:qubitConnectivity}, qubits $q_1$, $q_2$, and $q_3$ form a connected component; $q_1$ and $q_2$ may impact $q_3$, but $q_3$ is not directly connected to $q_1$.

In \QAICCC, crosstalk is quantified using a single
metrics like the \emph{composite score}, which is the sum of the
Stochastic and Hamiltonian error rates as defined in \citet{ASA+20}'s work.
Hence, for each \xTalk{m}{n} crosstalk, we obtain a triple $\dataList{\xTalkScore, \xTalkImpacting, \xTalkImpacted}$, where $\xTalkScore$ is the composite score, $\xTalkImpacting$ a set of $m$ impacting qubits, and $\xTalkImpacted$ a set of $n$ impacted qubits.
We denote by $\xTalkScores$ the list of such triples. For instance, the analysis of the platform shown in
~\Cref{fig:qubitConnectivity} using the \pyGSTi simulator, yields
the following composite scores (as reported by \citet{ASA+20} in their first run): $0.0027$ from $q_3$ and $q_4$ to $q_2$, $0.0024$ from $q_2$ and $q_4$ to $q_3$, etc.
Thus, $\xTalkScores = [ \dataList{0.0027, \set{q_3, q_4}{}, \set{q_2}{}}$, $\dataList{0.0024, \set{q_2, q_4}{}, \set{q_3}{}}$, $\dots]$.

 \subsection{Step 2: Allocation algorithm}
\label{sec:allocation}

In this step, we allocate qubits to users in order to meet G1 and G2,
using \Cref{algo:allocation}. The algorithm takes as input the
connectivity of the platform $C$, the information on the size  of the
user circuits $\Sizes$, and the list $\xTalkScores$ returned by Step~1 (\Cref{sec:crosstalk}).

\begin{algorithm}[t]
\caption{Qubit allocation algorithm}
\label{algo:allocation}
\footnotesize
\begin{algorithmic}[1]
\AlgoType $\typeAlloc = \dataList{\Unallocated, \Trusted, \Untrusted}$, where $\Unallocated$ is a set of qubits and $\Trusted$ and $\Untrusted$ are sets of sets of qubits
\AlgoInput $C = \dataList{V, E}$, where $V$ is a set of qubits and $E$ is a set of unordered pairs of qubits
\AlgoInput $\Sizes = \dataList{\SizeTrusted, \SizeUntrusted}$, where $\SizeTrusted$ and $\SizeUntrusted$ are sets of integers
\AlgoInput $\xTalkScores$: List of triples $\dataList{\xTalkScore, \xTalkImpacting, \xTalkImpacted}$ as in Step~1 (\Cref{sec:crosstalk})
\AlgoOutput Set of \typeAllocs or $\algoError$
\If{$\sum_{s \in \SizeTrusted} s + \sum_{s \in \SizeUntrusted} s > \card{V}$} \label{line:allocation:tooFew}
        \State \Return $\algoError$ \label{line:allocation:error} \Comment{Not enough qubits}
    \EndIf
    \State $\Sizes \leftarrow \updSizes{V, \Sizes}$\label{line:allocation:callUpdSizes}\Comment{Idle user, if needed}
    \State \typeAlloc $\allocation_0 \leftarrow \initAlloc{V, \xTalkScores}$ \label{line:allocation:initAlloc} 
    \State Set of \typeAllocs $\Population \leftarrow \set{\allocation_0}{}$ \label{line:allocation:initPopulation} \Comment{Initial population}
    \State Set of \typeAllocs $\Candidates \leftarrow \varnothing$ \label{line:allocation:initCandidates}
    \State $\xTalkScores \leftarrow \mathit{sort}(\xTalkScores, \mathit{reverse} = \algoTrue)$ \label{line:allocation:sort}
    \For{$\xTalkTriple \in \xTalkScores$} \Comment{Largest to smallest score} \label{line:allocation:forScore}
        \State Set of \typeAllocs $\CurrentPop \leftarrow \Population$ \label{line:allocation:currentPop}
        \For{$\allocation_1 \in \CurrentPop$} \label{line:allocation:iterAlloc}
            \State Set of \typeAllocs $\Allocations \leftarrow \varnothing$ \label{line:allocation:allocs}
            \If{$\isFixed{\allocation_1}{\xTalkTriple}$} \label{line:allocation:isFixed}\Comment{Safe allocation pattern}
                \If{$\nbUsers{\allocation_1}{\xTalkTriple} \ge 2$}\label{line:allocation:testUsers}
                    \State $\Allocations \leftarrow \improveAlloc{\allocation_1, \xTalkTriple, E, \Sizes}$ \label{line:allocation:severalUsers}
                \EndIf
                \State $\evalAlloc{\allocation_1}{\xTalkTriple}$ \label{line:allocation:evalCurrentAlloc} \Comment{Update attributes}
            \Else
                \State $\Allocations \leftarrow \allocUnallocated{\allocation_1, \xTalkImpacted, C, \Sizes}$ \label{line:allocation:allocUnallocated}
                \State $\Allocations \leftarrow \allocImpacted{\Allocations, \xTalkTriple, C, \Sizes}$ \label{line:allocation:allocImpacted}
                \State $\Allocations \leftarrow \Allocations \cup \improveAlloc{\allocation_1, \xTalkTriple, C, \Sizes}$ \label{line:allocation:mergeAllocs}
                \State $\Allocations \leftarrow \Allocations \cup \allocTrusted{\allocation_1, \xTalkImpacting, C, \Sizes}$ \label{line:allocation:allocTrusted}
                \State $\Population, \Candidates \leftarrow \archiveAlloc{\allocation_1, \Population, \Candidates}$ \label{line:allocation:archiveAlloc}
            \EndIf
                \State $\Population \leftarrow \updPop{\Allocations, \Population, \Candidates, \xTalkTriple}$ \label{line:allocation:updPop}
        \EndFor
        \If{$\Population = \varnothing$} \label{line:allocation:emptyPop}\Comment{No new change}
            \State \textbf{break} \label{line:allocation:break}
        \EndIf 
    \EndFor \label{line:allocation:endForScore}
    \State \Return $\Population \cup \Candidates$ \label{line:allocation:output}\Comment{All the obtained allocations}
\end{algorithmic}
 \end{algorithm}

The platform connectivity $C$ is represented as an undirected graph $\dataList{V,
  E}$, where the set of vertices $V$ corresponds to the qubits and the
set of edges $E$ corresponds to the connected qubit pairs.
As for the information on the size  of the
user circuit (i.e., the number of qubits
they require), since some users are trusted, we represent $\Sizes$ as the pair $\Sizes = \dataList{\SizeTrusted, \SizeUntrusted}$, where $\SizeTrusted$ and $\SizeUntrusted$ are sets of integers, respectively corresponding to the circuit sizes requested by trusted and untrusted users.

We introduce the concept of qubit \emph{allocation}, i.e., how the qubits in $V$ are either unallocated or allocated to users.
Each allocation, corresponding to the type \typeAlloc declared on the first line of \Cref{algo:allocation}, is
represented as a triple $\dataList{\Unallocated, \Trusted, \Untrusted}$, where $\Unallocated$ is the set of unallocated qubits and $\Trusted$ and $\Untrusted$ are sets of sets of qubits, representing the qubits allocated to, respectively, trusted and untrusted users.
Moreover, each allocation has several attributes:
\emph{\allocIncidentalName} is a list of remaining inter-circuit
\emph{incidental crosstalk},
(eventually) to be minimized in Step 4 (\Cref{sec:noiseReduction});
\emph{\allocScoreName} is the score used to rank allocations (the lower the score, the more secure the allocation);
\emph{\allocPenaltyName} represents the penalty of the
allocation, i.e., it is a metric quantifying cumulative incidental crosstalk and
used to rank allocations in case of a tie in score;
\emph{\xTalkLastName} is the first encountered error rate for which the allocation is unsafe.
For a given allocation, if a qubit is allocated to a user, we say that this user \emph{controls} the qubit.

The algorithm takes inspiration from genetic search by maintaining an evolving population of possible allocations, but does not involve any randomness.
To minimize the largest inter-circuit crosstalk error rate, it
processes error rates returned by Step~1 (\Cref{sec:crosstalk}) in
decreasing order, so that the qubits involved in crosstalk with the largest intensity are allocated first.
For a given error rate, each allocation in the population is tested to
determine if it is already safe, i.e., impacting qubits are allocated
to trusted users or impacted users control impacting qubits.
If the allocation is safe, then it is used in the next iteration.
Otherwise, it is removed from the population, and then (when possible,
due to size constraints) new safe allocations are generated from it and added to the population.
In this way, the last individuals in the population correspond to
allocations of qubits done in the safest possible way.

The algorithm starts by testing if the number of qubits to allocate is larger than the number of available qubits (\Cref{line:allocation:tooFew}).
If so, then no qubit allocation is possible and \QAICCC returns an error (\Cref{line:allocation:error}).
Instead, if the number of qubits to allocate is smaller than the number of available qubits, then this means some qubits are not used by the users' circuits.
In this case, to meet G1, the algorithm should allocate qubits for potential, future users.
To do so, we use the auxiliary function \texttt{\updSizesName}, which works as follows.
It introduces an untrusted \emph{idle user} requesting for
the rest of the qubits, so that all the qubits will be allocated, then it
updates the tuple $\Sizes$ accordingly (\Cref{line:allocation:callUpdSizes}).
Since each user's allocation will form a connected component, the unused qubits will be connected, which increases the chance of being able to allocate them to new users.
If the users already requested all the available qubits, then there is no need for an idle user and, thus, $\Sizes$ is not updated.

The initial allocation $\allocation_0$ is generated at
\Cref{line:allocation:initAlloc} such that all the qubits are
unallocated, its \emph{\allocScoreName} is higher than the ones in $\xTalkScores$ (to
reflect that it is unsafe), its \emph{\allocPenaltyName} is zero, and its
\emph{\allocIncidentalName} is empty.
The \emph{population}, denoted by $\Population$, is a set of
allocations which are safe for the crosstalk error rates in
$\xTalkScores$ investigated so far.
The population is initialized at \Cref{line:allocation:initPopulation} with the initial allocation.
The \emph{archive}, denoted by $\Candidates$, is a set of allocations
which are unsafe for at least one error rate; it is initialized empty at \Cref{line:allocation:initCandidates}.

To meet G2, we minimize the largest inter-circuit crosstalk error
rate.
The list of error rates $\xTalkScores$ obtained during crosstalk
analysis (\Cref{sec:crosstalk}) is first sorted in decreasing
order of $\xTalkScore$ (\Cref{line:allocation:sort});
then, each
error rate $\xTalkTriple = \dataList{\xTalkScore, \xTalkImpacting,
  \xTalkImpacted}$
is processed to obtain safe allocations
(\Crefrange{line:allocation:forScore}{line:allocation:endForScore}).
More specifically, at each iteration, the current population is stored in an ancillary variable (\Cref{line:allocation:currentPop}) and each allocation $\allocation_1$ in the current population is used in the body of the loop to (eventually) update $\Population$ (\Cref{line:allocation:iterAlloc}).
Depending on $\allocation_1$'s properties, new allocations are (eventually) generated and added to the population.
This set of new allocations is denoted by $\Allocations$ and initialized empty at \Cref{line:allocation:allocs}.
 Each new allocation inherits its attributes from $\allocation_1$ and is generated by updating $\allocation_1$ to address the current error rate.
Moreover, to ensure user circuits can be transpiled using the obtained qubit allocations, each user's allocation is determined so that it forms a connected component.
For instance, if two users respectively submit a $2$-qubit and a $3$-qubit circuit for the platform illustrated in \Cref{fig:qubitConnectivity}, an allocation where $q_0$, $q_2$, and $q_3$ are controlled by the same user will not be considered, as it is not possible to use $q_0$, $q_2$, and $q_3$ in one circuit and $q_1$ and $q_4$ in another one, since $q_1$ and $q_4$ are not connected.

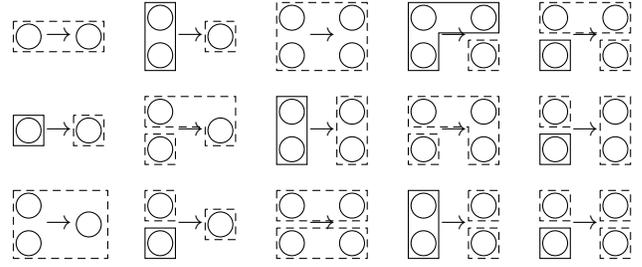
\begin{figure}[t]
\centering
\begin{tikzpicture}[scale=1.0]
\tikzset{
        qubit/.style={draw,circle},
        trustedUser/.style={color=black},
        anyUser/.style={color=black, densely dashed}
    }
\node[] (1A)
    	at (0,0*\ySpace)
	{$\rightarrow$};
    \node[qubit] (1B)
    	at ($(1A)+(-\xBetween,0)$)
	{};
    \node[qubit] (1C)
    	at ($(1A)+(\xBetween,0)$)
	{};
    \draw[anyUser]
    	($(1B.north west)+(-\offsetBox,\offsetBox)$)
	rectangle
	($(1C.south east)+(\offsetBox,-\offsetBox)$);
\node[] (2A)
    	at (0,-1*\ySpace)
	{$\rightarrow$};
    \node[qubit] (2B)
    	at ($(2A)+(-\xBetween,0)$)
	{};
    \node[qubit] (2C)
    	at ($(2A)+(\xBetween,0)$)
	{};
    \draw[trustedUser]
    	($(2B.north west)+(-\offsetBox,\offsetBox)$)
	rectangle
	($(2B.south east)+(\offsetBox,-\offsetBox)$);
    \draw[anyUser]
    	($(2C.north west)+(-\offsetBox,\offsetBox)$)
	rectangle
	($(2C.south east)+(\offsetBox,-\offsetBox)$);
\node[] (3A)
    	at (0,-2*\ySpace)
	{$\rightarrow$};
    \node[qubit] (3B)
    	at ($(3A)+(-\xBetween,\offsetQubit)$)
	{};
    \node[qubit] (3C)
    	at ($(3A)+(-\xBetween,-\offsetQubit)$)
	{};
    \node[qubit] (3D)
    	at ($(3A)+(\xBetween,0)$)
	{};
    \ExtractCoordinateA{3C.south west}
    \ExtractCoordinateB{3D.east}
    \draw[anyUser]
    	($(3B.north west)+(-\offsetBox,\offsetBox)$)
	rectangle
	($(\XCoordB, \YCoordA)+(\offsetBox,-\offsetBox)$);
\node[] (4A)
    	at (\xSpace,0)
	{$\rightarrow$};
    \node[qubit] (4B)
    	at ($(4A)+(-\xBetween,\offsetQubit)$)
	{};
    \node[qubit] (4C)
    	at ($(4A)+(-\xBetween,-\offsetQubit)$)
	{};
    \node[qubit] (4D)
    	at ($(4A)+(\xBetween,0)$)
	{};
    \draw[trustedUser]
    	($(4B.north west)+(-\offsetBox,\offsetBox)$)
	rectangle
	($(4C.south east)+(\offsetBox,-\offsetBox)$);
    \draw[anyUser]
    	($(4D.north west)+(-\offsetBox,\offsetBox)$)
	rectangle
	($(4D.south east)+(\offsetBox,-\offsetBox)$);
\node[] (5A)
    	at (\xSpace,-\ySpace)
	{$\rightarrow$};
    \node[qubit] (5B)
    	at ($(5A)+(-\xBetween,\offsetQubit)$)
	{};
    \node[qubit] (5C)
    	at ($(5A)+(-\xBetween,-\offsetQubit)$)
	{};
    \node[qubit] (5D)
    	at ($(5A)+(\xBetween,0)$)
	{};
    \ExtractCoordinateA{5B.north east}
    \ExtractCoordinateB{5D.north east}
    \coordinate (corner5A) at (\XCoordB, \YCoordA);
    \ExtractCoordinateA{5B.south west}
    \ExtractCoordinateB{5D.north west}
    \coordinate (corner5B) at (\XCoordB, \YCoordA);
    \draw[anyUser]
    	($(5B.south west)+(-\offsetBox,-\offsetBox)$) --
    	($(5B.north west)+(-\offsetBox,\offsetBox)$) --
    	($(corner5A)+(\offsetBox,\offsetBox)$) --
    	($(5D.south east)+(\offsetBox,-\offsetBox)$) --
    	($(5D.south west)+(-\offsetBox,-\offsetBox)$) --
    	($(corner5B)+(-\offsetBox,-\offsetBox)$) --
    	cycle;
    \draw[anyUser]
    	($(5C.north west)+(-\offsetBox,\offsetBox)$)
	rectangle
	($(5C.south east)+(\offsetBox,-\offsetBox)$);
\node[] (6A)
    	at (\xSpace,-2*\ySpace)
	{$\rightarrow$};
    \node[qubit] (6B)
    	at ($(6A)+(-\xBetween,\offsetQubit)$)
	{};
    \node[qubit] (6C)
    	at ($(6A)+(-\xBetween,-\offsetQubit)$)
	{};
    \node[qubit] (6D)
    	at ($(6A)+(\xBetween,0)$)
	{};
    \draw[anyUser]
    	($(6B.north west)+(-\offsetBox,\offsetBox)$)
	rectangle
	($(6B.south east)+(\offsetBox,-\offsetBox)$);
    \draw[trustedUser]
    	($(6C.north west)+(-\offsetBox,\offsetBox)$)
	rectangle
	($(6C.south east)+(\offsetBox,-\offsetBox)$);
    \draw[anyUser]
    	($(6D.north west)+(-\offsetBox,\offsetBox)$)
	rectangle
	($(6D.south east)+(\offsetBox,-\offsetBox)$);
\node[] (7A)
    	at (2*\xSpace,0)
	{$\rightarrow$};
    \node[qubit] (7B)
    	at ($(7A)+(-\xBetween,\offsetQubit)$)
	{};
    \node[qubit] (7C)
    	at ($(7A)+(-\xBetween,-\offsetQubit)$)
	{};
    \node[qubit] (7D)
    	at ($(7A)+(\xBetween,\offsetQubit)$)
	{};
    \node[qubit] (7E)
    	at ($(7A)+(\xBetween,-\offsetQubit)$)
	{};
    \draw[anyUser]
    	($(7B.north west)+(-\offsetBox,\offsetBox)$)
	rectangle
    	($(7E.south east)+(\offsetBox,-\offsetBox)$);
\node[] (8A)
    	at (2*\xSpace,-\ySpace)
	{$\rightarrow$};
    \node[qubit] (8B)
    	at ($(8A)+(-\xBetween,\offsetQubit)$)
	{};
    \node[qubit] (8C)
    	at ($(8A)+(-\xBetween,-\offsetQubit)$)
	{};
    \node[qubit] (8D)
    	at ($(8A)+(\xBetween,\offsetQubit)$)
	{};
    \node[qubit] (8E)
    	at ($(8A)+(\xBetween,-\offsetQubit)$)
	{};
    \draw[trustedUser]
    	($(8B.north west)+(-\offsetBox,\offsetBox)$)
	rectangle
    	($(8C.south east)+(\offsetBox,-\offsetBox)$);
    \draw[anyUser]
    	($(8D.north west)+(-\offsetBox,\offsetBox)$)
	rectangle
    	($(8E.south east)+(\offsetBox,-\offsetBox)$);
\node[] (9A)
    	at (2*\xSpace,-2*\ySpace)
	{$\rightarrow$};
    \node[qubit] (9B)
    	at ($(9A)+(-\xBetween,\offsetQubit)$)
	{};
    \node[qubit] (9C)
    	at ($(9A)+(-\xBetween,-\offsetQubit)$)
	{};
    \node[qubit] (9D)
    	at ($(9A)+(\xBetween,\offsetQubit)$)
	{};
    \node[qubit] (9E)
    	at ($(9A)+(\xBetween,-\offsetQubit)$)
	{};
    \draw[anyUser]
    	($(9B.north west)+(-\offsetBox,\offsetBox)$)
	rectangle
    	($(9D.south east)+(\offsetBox,-\offsetBox)$);
    \draw[anyUser]
    	($(9C.north west)+(-\offsetBox,\offsetBox)$)
	rectangle
    	($(9E.south east)+(\offsetBox,-\offsetBox)$);
\node[] (10A)
    	at (3*\xSpace,0)
	{$\rightarrow$};
    \node[qubit] (10B)
    	at ($(10A)+(-\xBetween,\offsetQubit)$)
	{};
    \node[qubit] (10C)
    	at ($(10A)+(-\xBetween,-\offsetQubit)$)
	{};
    \node[qubit] (10D)
    	at ($(10A)+(\xBetween,\offsetQubit)$)
	{};
    \node[qubit] (10E)
    	at ($(10A)+(\xBetween,-\offsetQubit)$)
	{};
    \ExtractCoordinateA{10D.south east}
    \ExtractCoordinateB{10C.south east}
    \coordinate (corner10) at (\XCoordB, \YCoordA);
    \draw[trustedUser]
    	($(10C.south east)+(\offsetBox,-\offsetBox)$) --
    	($(10C.south west)+(-\offsetBox,-\offsetBox)$) --
    	($(10B.north west)+(-\offsetBox,\offsetBox)$) --
    	($(10D.north east)+(\offsetBox,\offsetBox)$) --
    	($(10D.south east)+(\offsetBox,-\offsetBox)$) --
    	($(corner10)+(\offsetBox,-\offsetBox)$) --
    	cycle;
    \draw[anyUser]
    	($(10E.north west)+(-\offsetBox,\offsetBox)$)
	rectangle
    	($(10E.south east)+(\offsetBox,-\offsetBox)$);
\node[] (11A)
    	at (3*\xSpace,-\ySpace)
	{$\rightarrow$};
    \node[qubit] (11B)
    	at ($(11A)+(-\xBetween,\offsetQubit)$)
	{};
    \node[qubit] (11C)
    	at ($(11A)+(-\xBetween,-\offsetQubit)$)
	{};
    \node[qubit] (11D)
    	at ($(11A)+(\xBetween,\offsetQubit)$)
	{};
    \node[qubit] (11E)
    	at ($(11A)+(\xBetween,-\offsetQubit)$)
	{};
    \ExtractCoordinateA{11B.south west}
    \ExtractCoordinateB{11E.south west}
    \coordinate (corner11) at (\XCoordB, \YCoordA);
    \draw[anyUser]
    	($(11B.south west)+(-\offsetBox,-\offsetBox)$) --
    	($(11B.north west)+(-\offsetBox,\offsetBox)$) --
    	($(11D.north east)+(\offsetBox,\offsetBox)$) --
    	($(11E.south east)+(\offsetBox,-\offsetBox)$) --
    	($(11E.south west)+(-\offsetBox,-\offsetBox)$) --
    	($(corner11)+(-\offsetBox,-\offsetBox)$) --
    	cycle;
    \draw[anyUser]
    	($(11C.north west)+(-\offsetBox,\offsetBox)$)
	rectangle
    	($(11C.south east)+(\offsetBox,-\offsetBox)$);
\node[] (12A)
    	at (3*\xSpace,-2*\ySpace)
	{$\rightarrow$};
    \node[qubit] (12B)
    	at ($(12A)+(-\xBetween,\offsetQubit)$)
	{};
    \node[qubit] (12C)
    	at ($(12A)+(-\xBetween,-\offsetQubit)$)
	{};
    \node[qubit] (12D)
    	at ($(12A)+(\xBetween,\offsetQubit)$)
	{};
    \node[qubit] (12E)
    	at ($(12A)+(\xBetween,-\offsetQubit)$)
	{};
    \draw[trustedUser]
    	($(12B.north west)+(-\offsetBox,\offsetBox)$)
	rectangle
    	($(12C.south east)+(\offsetBox,-\offsetBox)$);
    \draw[anyUser]
    	($(12D.north west)+(-\offsetBox,\offsetBox)$)
	rectangle
    	($(12D.south east)+(\offsetBox,-\offsetBox)$);
    \draw[anyUser]
    	($(12E.north west)+(-\offsetBox,\offsetBox)$)
	rectangle
    	($(12E.south east)+(\offsetBox,-\offsetBox)$);
\node[] (13A)
    	at (4*\xSpace,0)
	{$\rightarrow$};
    \node[qubit] (13B)
    	at ($(13A)+(-\xBetween,\offsetQubit)$)
	{};
    \node[qubit] (13C)
    	at ($(13A)+(-\xBetween,-\offsetQubit)$)
	{};
    \node[qubit] (13D)
    	at ($(13A)+(\xBetween,\offsetQubit)$)
	{};
    \node[qubit] (13E)
    	at ($(13A)+(\xBetween,-\offsetQubit)$)
	{};
    \draw[anyUser]
    	($(13B.north west)+(-\offsetBox,\offsetBox)$)
	rectangle
    	($(13D.south east)+(\offsetBox,-\offsetBox)$);
    \draw[trustedUser]
    	($(13C.north west)+(-\offsetBox,\offsetBox)$)
	rectangle
    	($(13C.south east)+(\offsetBox,-\offsetBox)$);
    \draw[anyUser]
    	($(13E.north west)+(-\offsetBox,\offsetBox)$)
	rectangle
    	($(13E.south east)+(\offsetBox,-\offsetBox)$);
\node[] (14A)
    	at (4*\xSpace,-\ySpace)
	{$\rightarrow$};
    \node[qubit] (14B)
    	at ($(14A)+(-\xBetween,\offsetQubit)$)
	{};
    \node[qubit] (14C)
    	at ($(14A)+(-\xBetween,-\offsetQubit)$)
	{};
    \node[qubit] (14D)
    	at ($(14A)+(\xBetween,\offsetQubit)$)
	{};
    \node[qubit] (14E)
    	at ($(14A)+(\xBetween,-\offsetQubit)$)
	{};
    \draw[anyUser]
    	($(14B.north west)+(-\offsetBox,\offsetBox)$)
	rectangle
    	($(14B.south east)+(\offsetBox,-\offsetBox)$);
    \draw[trustedUser]
    	($(14C.north west)+(-\offsetBox,\offsetBox)$)
	rectangle
    	($(14C.south east)+(\offsetBox,-\offsetBox)$);
    \draw[anyUser]
    	($(14D.north west)+(-\offsetBox,\offsetBox)$)
	rectangle
    	($(14E.south east)+(\offsetBox,-\offsetBox)$);
\node[] (15A)
    	at (4*\xSpace,-2*\ySpace)
	{$\rightarrow$};
    \node[qubit] (15B)
    	at ($(15A)+(-\xBetween,\offsetQubit)$)
	{};
    \node[qubit] (15C)
    	at ($(15A)+(-\xBetween,-\offsetQubit)$)
	{};
    \node[qubit] (15D)
    	at ($(15A)+(\xBetween,\offsetQubit)$)
	{};
    \node[qubit] (15E)
    	at ($(15A)+(\xBetween,-\offsetQubit)$)
	{};
    \draw[anyUser]
    	($(15B.north west)+(-\offsetBox,\offsetBox)$)
	rectangle
    	($(15B.south east)+(\offsetBox,-\offsetBox)$);
    \draw[trustedUser]
    	($(15C.north west)+(-\offsetBox,\offsetBox)$)
	rectangle
    	($(15C.south east)+(\offsetBox,-\offsetBox)$);
    \draw[anyUser]
    	($(15D.north west)+(-\offsetBox,\offsetBox)$)
	rectangle
    	($(15D.south east)+(\offsetBox,-\offsetBox)$);
    \draw[anyUser]
    	($(15E.north west)+(-\offsetBox,\offsetBox)$)
	rectangle
    	($(15E.south east)+(\offsetBox,-\offsetBox)$);
\end{tikzpicture}
 \caption{Safe allocation patterns for \xTalk{1}{1}, \xTalk{2}{1}, and \xTalk{2}{2} crosstalk: impacting qubits $\xTalkImpacting$ are on the left of an arrow and impacted qubits $\xTalkImpacted$ on the right, each box with plain (resp. dashed) edges represents qubits allocated to a trusted (resp. any) user.}
\label{fig:safePatterns}
\end{figure}

First, $\allocation_1$ is tested to determine if it is already safe
with respect to $\xTalkTriple$ (\Cref{line:allocation:isFixed}), i.e.,
if the allocation matches a \emph{safe allocation pattern} depicted in \Cref{fig:safePatterns}.
In short, a pattern is safe if a trusted user controls at least one impacting qubit or if all impacted users control at least one impacting qubit\footnote{It is not enough to spread the control of impacting qubits to different users, since they may be in collusion or simply be the same actor having submitted several circuits.}.
Note that trusted users are not necessary, as several safe patterns for \xTalk{1}{1}, \xTalk{2}{1}, and \xTalk{2}{2} crosstalk do not involve them.

Second, in case qubits involved in crosstalk are allocated to several users, some crosstalk may occur when these qubits are activated at the same time.
In the context of this paper, this phenomenon is called \emph{incidental crosstalk} since, as opposed to a crosstalk attack, it is not intentionally triggered by a user.
Thus, even if the allocation matches a safe pattern, it is tested to determine the number of users controlling the qubits of $\xTalkTriple$ (\Cref{line:allocation:testUsers}).
If only one user is involved, then there is no need for new allocations, thus $\Allocations$ remains empty.
Otherwise, new allocations involving only one user are generated using
the auxiliary function
\texttt{\improveAllocName} (\Cref{line:allocation:severalUsers}).
The latter works as follows: 
if all the involved qubits are unallocated, then they will; either allocated to a user controlling no qubit (if possible) or connected to qubits of users in a way that does not violate the circuit size limit.
Otherwise, if some involved qubits are allocated, then they will be reallocated to a unique user by merging the allocations of users controlling an involved qubit.
At this point (\Cref{line:allocation:evalCurrentAlloc}), since
$\allocation_1$ is safe, it is kept for the next iteration and its
attributes are updated using the auxiliary function \texttt{\evalAllocName} as follows:
its score becomes $\xTalkScore$ and, if $\allocation_1$ involves several users, then its penalty is increased by $\xTalkScore$ and $\xTalkTriple$ is appended to $\allocIncidental{\allocation_1}$.

If $\allocation_1$ is unsafe, then new safe allocations are generated
and stored in $\Allocations$ as follows (\Crefrange{line:allocation:allocUnallocated}{line:allocation:allocTrusted}).
First, all the unallocated impacted qubits are allocated to users
(\Cref{line:allocation:allocUnallocated}); then, function
\texttt{\allocImpactedName} updates the allocations so that each impacted user
$u_2$ controls an impacting qubit (\Cref{line:allocation:allocImpacted}).
These updates are achieved either by allocating an unallocated (if
any) impacting qubit to $u_2$ or by merging $u_2$'s qubits with qubits
controlled by a user $u_1$ controlling an impacting qubit.
At this stage, the set of allocation is further updated, by
allocating all the qubits involved in the error rate  
to the same user (\Cref{line:allocation:mergeAllocs})---as in
\Cref{line:allocation:severalUsers}---and  by allocating an
unallocated (if any) impacting qubit to a trusted user
(\Cref{line:allocation:allocTrusted}).
Then, since $\allocation_1$ is unsafe, it has to be removed from the
current population. This is achieved by calling function \texttt{\archiveAllocName},
which sets $\xTalkLast{\allocation_1}$ to $\xTalkTriple$, removes
$\allocation_1$  from $\Population$, and adds it to $\Candidates$.
Since only allocations in the population will be investigated in future iterations, $\allocation_1$'s attributes will not change until the end of the algorithm.

After these tests, $\Allocations$ (if not empty) contains allocations which are safe for all the crosstalk error rates investigated so far.
The population is (eventually) updated at \Cref{line:allocation:updPop}.
More precisely, for each $\allocation_2$ in $\Allocations$, if $\allocation_2$ does not already belong to $\Population \cup \Candidates$, then its attributes are updated (as in \Cref{line:allocation:evalCurrentAlloc}) and it is added to the population.
The purpose of this condition is to prevent a conflict of attributes between several occurrences of the same allocation.

Finally, if no allocation
can address $\xTalkTriple$, then the population becomes empty (\Cref{line:allocation:emptyPop}).
Since no population change can occur from this state, the loop stops (\Cref{line:allocation:break}).
In this case or if all crosstalk error rates were exhausted, \Cref{algo:allocation} returns all the obtained allocations (\Cref{line:allocation:output}).

The additional auxiliary functions used in \Cref{algo:allocation} are
detailed 
\ifreport
in Appendix~\ref{sec:auxiliaryFunctions2}.
\else
in the appendix of the online version of the paper~\cite{MB25}.
\fi

 \subsection{Step 3: Transpilation}
\label{sec:transpilation}

In this step, user circuits are transpiled according to the selected qubit allocation.

Allocations returned by \Cref{algo:allocation} are first sorted by
increasing $\allocScoreSymb$;  in case of a tie, they are further
sorted by increasing $\allocPenaltySymb$
In this way, the initial (empty) allocation is the last one, while the
allocations obtained last with \Cref{algo:allocation} (i.e., those addressing the largest number of error rates) are the first ones.
Starting from the first allocation, an allocation is selected and tested as follows.
\Qiskit is called to transpile the user circuits according to the current allocation, to match the platform connectivity and supported gates.
If the transpilation is successful, then the transpiled circuits are selected for Step 4.
Otherwise, the next allocation is selected and tested.
This process continues until an allocation is successful.
If no allocation is successful, then \QAICCC returns an error.

During transpilation, non-binary quantum gates are decomposed in binary gates supported by the platform; hence, a transpiled circuit will contain only binary quantum gates.
Moreover, since qubits allocated to each circuit form a connected component
(\Cref{sec:allocation}) and swap gates ensure that, in a connected
component, any pair of qubits can be used in a quantum operation, it
is likely that all allocations would be successfully transpiled
(depending on Qiskit's behavior).

 \subsection{Step 4: Noise reduction}
\label{sec:noiseReduction}

In this step, \QAICCC uses existing techniques like the \XtalkSched
scheduler~\cite{MMM+20} and \ColorDynamic~\cite{DGL+20}
(\Cref{sec:background}) on the transpiled circuits to reduce noise
further.
These techniques can be used either in isolation or in combination.
\ColorDynamic can be used only if the platform supports
frequency allocation, while there are  no preconditions for
\XtalkSched.
If used in combination, since \ColorDynamic  does not impact the
circuit schedule, it should be used first (to reduce crosstalk) and
then followed by  \XtalkSched (to reduce the remaining noise).

To meet G2, the priority is to reduce first remaining inter-circuit
crosstalk of the qubit allocation selected in Step~3
(\Cref{sec:transpilation}), i.e., error rates in its attribute
\emph{\allocIncidentalName};  then, error rates starting from its
attribute \emph{\xTalkLastName} are also considered as candidates for
reduction.
Note that these error rates are already sorted by decreasing order of $\xTalkScore$ (\Cref{sec:allocation}).
Hence, noise reduction techniques can be applied following the same order to reduce the error rates.

To meet G3, the aforementioned noise reduction techniques can be also independently applied to each circuit, to reduce intra-circuit noise (crosstalk and decoherence).
Since G2 has higher priority than G3, such techniques have to be
applied in a way that preserves the changes (e.g., frequency allocation for \ColorDynamic and gate scheduling for \XtalkSched) required to meet G2.

\subsection{Application to the Running Example}

Let us assume that two untrusted users respectively submit a $2$-qubit and a $3$-qubit circuit for the platform illustrated in \Cref{fig:qubitConnectivity}.
Step~1 obtains $\xTalkScores = [$
$\dataList{0.0027, \set{q_3, q_4}{}, \set{q_2}{}}$,
$\dataList{0.0017, \set{q_1, q_2}{}, \set{q_0}{}}$,
$\dataList{0.0013, \set{q_2, q_4}{}, \set{q_0}{}}$,
$\dots]$.

In Step 2, the allocation algorithm determines that all the qubits are used, hence no idle user is necessary (\Cref{line:allocation:callUpdSizes}).
The population is initialized (\Cref{line:allocation:initPopulation}) with the initial individual
$\dataList{\Unallocated, \Trusted, \Untrusted}$
where all the qubits are unallocated, i.e., $\Unallocated = \set{q_0, q_1, q_2, q_3, q_4}{}$, $\Trusted = \varnothing$, and $\Untrusted = \varnothing$.
Since there is no trusted user in this example, we simply denote an allocation by $\dataList{\Untrusted}$, omitting $\Unallocated = E \setminus \Untrusted$ and $\Trusted = \varnothing$.
The iteration (\Cref{line:allocation:forScore}) starts with the largest error rate $\dataList{0.0027, \set{q_3, q_4}{}, \set{q_2}{}}$.
Since the only individual present in $\Population$ is $\dataList{\varnothing}$, which is unsafe since qubits involved in the error rate could be allocated to any user, it is archived and new allocations are generated.
First, the impacted qubit is allocated (\Cref{line:allocation:allocUnallocated}), obtaining the allocation $\dataList{\set{\set{q_2}{}}{}}$.
Second, impacting qubits are allocated to the user controlling the impacted qubit (\Cref{line:allocation:allocImpacted}), so $\dataList{\set{\set{q_2}{}}{}}$ is replaced by $\dataList{\set{\set{q_2, q_3}{}}{}}$ and $\dataList{\set{\set{q_2, q_4}{}}{}}$.
Because the unallocated controlling qubit could be allocated to another user, these allocations have a penalty of $0.0027$.
Then, the algorithm adds allocation $\dataList{\set{\set{q_2, q_3, q_4}{}}{}}$, where only one user is involved, to the population (\Cref{line:allocation:mergeAllocs}).
This allocation has the same score (i.e., $0.0027$) as the allocations $\dataList{\set{\set{q_2, q_3}{}}{}}$ and $\dataList{\set{\set{q_2, q_4}{}}{}}$, but its penalty is $0$.
Since no user is trusted in this example, \Cref{line:allocation:allocTrusted} never generates new allocations.
Then, the algorithm continues with the next error rate $\dataList{0.0017, \set{q_1, q_2}{}, \set{q_0}{}}$.
To address this error rate, the population evolves from
$\{\dataList{\set{\set{q_2, q_3}{}}{}}$,
$\dataList{\set{\set{q_2, q_4}{}}{}}$,
$\dataList{\set{\set{q_2, q_3, q_4}{}}{}}\}$
to
$\{\dataList{\set{\set{q_0, q_1}{},\set{q_2, q_3}{}}{}}$,
$\dataList{\set{\set{q_0, q_1}{},\set{q_2, q_4}{}}{}}$,
$\dataList{\set{\set{q_0, q_1}{}\set{q_2, q_3, q_4}{}}{}}\}$.
Finally, the algorithm continues with the next error rate $\dataList{0.0013, \set{q_2, q_4}{}, \set{q_0}{}}$.
This time, no allocation is safe and no safe allocation can be generated.
Hence, the allocations are archived (\Cref{line:allocation:archiveAlloc}) and the iteration stops (\Cref{line:allocation:break}).
Finally, \Cref{algo:allocation} returns the archived allocations (\Cref{line:allocation:output}).

In Step 3, \QAICCC will select allocation $\dataList{\set{\set{q_0, q_1}{}, \set{q_2, q_3, q_4}{}}{}}$, as it has the same score as the other best allocations, but has a smaller penalty.
Both circuits would successfully be transpiled according to this allocation, which reduces inter-circuit crosstalk compared to the allocation $\dataList{\set{\set{q_0, q_1, q_2}{}, \set{q_3, q_4}{}}{}}$ exploited by \citet{ASA+20}'s injection attack.

However, even if reduced, inter-circuit crosstalk may still subsist.
In Step 4, noise reduction techniques may be used to further reduce noise.
For instance, \XtalkSched can be used to reduce inter-circuit crosstalk,
e.g., from $q_2$ and $q_4$ to $q_0$,
then to reduce intra-circuit crosstalk,
e.g., from $q_2$ and $q_3$ to $q_4$.

 \section{Related Work}
\label{sec:related}

\citet{ASA+20} proposed buffer qubits between user circuits to prevent
inter-circuit crosstalk as well as the injection attack they demonstrated.
While such an approach can meet G2, it would be a waste of quantum resources, failing to meet G1.
This contrasts with \QAICCC, which does not need to unallocate qubits to
prevent attacks and  makes unused qubits accessible for the next users.

To the best of our knowledge, existing reducing-crosstalk techniques like \XtalkSched~\cite{MMM+20} and \ColorDynamic~\cite{DGL+20} do not take into account security vulnerabilities.
Nevertheless, they are useful to meet G3 and, if they can be updated to prioritize inter-circuit over intra-circuit crosstalk and larger error rates over smaller ones (\Cref{sec:noiseReduction}), then they can contribute to meeting G2.
Moreover, \ColorDynamic can only be used when qubit operating
frequencies can be controlled by the software (e.g., tunable qubit architectures proposed in some prototypes~\cite{HHL+17} or by Google~\cite{BQP+19}), while \XtalkSched and \QAICCC can be used on any platform.
Finally, \XtalkSched focuses on gate errors, i.e., \xTalk{2}{2} crosstalk (\Cref{sec:allocation}), while \QAICCC uses qubit allocation to also reduce \xTalk{1}{1} and \xTalk{2}{1} crosstalk.

 \section{Research Outlook and Conclusions}\label{sec:conclusion}

In this idea paper, we presented \QAICCC, an approach for securing users'
program executions from crosstalk attacks in quantum servers. We
proposed a qubit allocation algorithm that maximizes qubit usage while
minimizing the largest inter-circuit crosstalk error rate.
Below, we outline our research plan to further develop this idea into
a full-fledged solution.

First, we plan to set-up an environment for \Qiskit and \pyGSTi to
replicate \citet{ASA+20}'s crosstalk attack on several platforms.
This attack acts on qubits involved in crosstalk and consists of executing several times a CNOT gate on impacting qubit(s) to reduce the magnitude of the desired output in the impacted qubit(s) below the magnitude of other results.
Thus, for various platform and qubit combinations, we will be able to
determine the correlation between the number of CNOT gate executions
necessary for the attack and the different error rates measured by
\pyGSTi. Such a correlation will allow us to assess whether the composite score they proposed in their
study is the best surrogate metric to quantify the strength of the
threat;
shouldn't this be the case, we will investigate alternative metrics.

Second, we will implement \QAICCC in \Python for better integration
with frameworks like \Qiskit and \PennyLane, and apply it to various
platforms and quantum circuits.

Furthermore, we plan to conduct an empirical evaluation of \QAICCC (either with a simulator or with a quantum processor, depending of available ressources).
The qubit allocation algorithm will be evaluated by comparing the largest inter-circuit crosstalk error rate between its allocation and an allocation performed by the last version of \Qiskit.
\QAICCC will be evaluated by simulating various combinations of
attacker, victim, trusted, and untrusted users and comparing---in
terms of success rate---the
execution of \QAICCC's transpiled circuits with transpiled circuits
obtained by baselines like \Qiskit and \XtalkSched alone.

 \section*{Acknowledgment}
This project has received funding from SES and the Luxembourg National Research Fund under the Industrial Partnership Block Grant (IPBG), ref. IPBG19/14016225/INSTRUCT.

\bibliographystyle{./Bibliography/IEEEtranN}

\ifreport
\clearpage
\appendix
\subsection{Auxiliary functions}
\label{sec:auxiliaryFunctions2}

The additional auxiliary functions used in \Cref{algo:allocation} are detailed in \Cref{algo:aux3}.
Function \texttt{\allocUnallocatedName} allocates unallocated impacted qubits; it takes as input the current allocation $\allocation_1$ and the impacted qubits $\xTalkImpacted$,
and returns the set of allocations $\Allocations_1$.
Function \texttt{\allocImpactedName} updates allocations in $\Allocations_1$ so that each impacted user controls at least one impacting qubit; it takes as input the set of allocations $\Allocations_1$ and the current error rate $\xTalkTriple$,
and returns the updated set of allocations $\Allocations_1$.
Function \texttt{\improveAllocName} generates new allocations involving only one user; it takes as input the current allocation $\allocation_1$ and the current error rate $\xTalkTriple$,
and returns the set of allocations $\Allocations_1$.
Function \texttt{\allocTrustedName} allocates unallocated impacting qubits to trusted users; it takes as input the current allocation $\allocation_1$ and the impacting qubits $\xTalkImpacting$,
and returns the set of allocations $\Allocations$.

These four auxiliary functions also take as input (but without using them directly) the platform connectivity $C = \dataList{V, E}$ and the circuit sizes $\Sizes$ because they call the \texttt{\connectName} and \texttt{\newAllocName} functions, which require these inputs.
Function \texttt{\connectName} is responsible for finding possible paths to connect qubits in a set of qubits $\User$ with qubits in a set of qubits $S$ and for using these paths to obtain (if any) connected components matching the circuit size limits.
It takes as input an allocation $\allocation_1$, the sets of qubits $\User$ and $S$, the platform connectivity $C$ and the circuit sizes $\Sizes$; it returns a set of allocations $\Allocations$.
If $\User$ is empty (e.g., when \texttt{\connectName} is called but no qubit has been allocated so far), then paths consist of either $S$ itself, if it is connected, or connected components containing $S$.
Finally, \texttt{\newAllocName} generates a new allocation;
it takes as input an allocation $\allocation_1$, the set of qubits $\User$, the current error rate $\xTalkTriple$, the platform connectivity $C$ and the circuit sizes $\Sizes$, and returns a set of allocations $\Allocations$.
A new allocation $\allocation_2$ is generated based on $\allocation_1$, so qubits in $\User$ are allocated to the same user.
Then, \texttt{\newAllocName} checks if $\allocation_2$ is compatible with the platform connectivity $C$ and satisfies the circuit size limits $\Sizes$.
If so, $\allocation_2$'s attributes are updated by the auxiliary function \texttt{\evalAllocName} according to the error rate $\xTalkTriple$, then $\Allocations = \set{\allocation_2}{}$ is returned;
otherwise, $\Allocations = \varnothing$ is returned.

\algrenewcommand\algorithmicindent{0.49em}\begin{figure}[t]
\begin{algorithmic}[1]
\footnotesize
\Procedure{$\allocUnallocatedSymb$}{$\allocation_1, \xTalkImpacted, E, \Sizes$}
    \State Set of \typeAllocs $\Allocations_1 \leftarrow \set{\allocation_1}{}$
    \For{Qubit $q \in \xTalkImpacted$}
        \State Set of \typeAllocs $\Allocations_2 \leftarrow \varnothing$
        \For{\typeAlloc $\allocation_2 = \dataList{\Unallocated, \Trusted, \Untrusted} \in \Allocations_1$}
            \If{$q \in \Unallocated$}
                \For{Set of Qubits $\User \in \Trusted \cup \Untrusted$}
                    \State $\Allocations_2 \leftarrow \Allocations_2 \cup \connect{\allocation_2, \User, \set{q}{}, C, \Sizes}$
                \EndFor
            \Else
                \State $\Allocations_2 \leftarrow \Allocations_2 \cup \set{\allocation_2}{}$
            \EndIf
        \EndFor
        \State $\Allocations_1 \leftarrow \Allocations_2$
    \EndFor
    \State \Return $\Allocations_1$
\EndProcedure
\State
\Procedure{$\allocImpactedSymb$}{$\Allocations_1, \xTalkTriple = \dataList{\xTalkScore, \xTalkImpacting, \xTalkImpacted}, C, \Sizes$}
    \For{Qubit $q_1 \in \xTalkImpacted$}
        \State Set of \typeAllocs $\Allocations_2 \leftarrow \varnothing$
        \For{\typeAlloc $\allocation_2 = \dataList{\Unallocated, \Trusted, \Untrusted} \in \Allocations_1$}
            \State Set of Qubits $\User_1 \leftarrow \getUser(q_1, \allocation_2)$
            \For{Qubit $q_2 \in \xTalkImpacting$}
                \If{$q_2 \in \Unallocated$}
                    \State $\Allocations_2 \leftarrow \Allocations_2 \cup \connect{\allocation_2, \User_1, \set{q_2}{}, C, \Sizes}$
                \Else
                    \State Set of Qubits $\User_2 \leftarrow \getUser(q_2, \allocation_2)$
                    \State $\Allocations_2 \leftarrow \Allocations_2 \cup \connect{\allocation_2, \User_1, \User_2, C, \Sizes}$
                \EndIf
            \EndFor
        \EndFor
        \State $\Allocations_1 \leftarrow \Allocations_2$
    \EndFor
    \State \Return $\Allocations_1$
\EndProcedure
\State
\Procedure{$\improveAllocSymb$}{$\allocation_1 = \dataList{\Unallocated, \Trusted, \Untrusted}, \xTalkTriple = \dataList{\xTalkScore, \xTalkImpacting, \xTalkImpacted}, C, \Sizes$}
    \State Set of \typeAllocs $\Allocations_1 \leftarrow \varnothing$
    \State Set of Qubits $S \leftarrow \xTalkImpacting \cup \xTalkImpacted$ \Comment{involved qubits}
    \State Set of Qubits $\UserMerge \leftarrow \bigcup_{q \in S} \getUser(q, \allocation_1)$
    \If{$\UserMerge = \varnothing$} \Comment{are unallocated}
        \State Set of \typeAllocs $\Allocations_2 \leftarrow \newAlloc(\allocation_1, S, \xTalkTriple, C, \Sizes)$
        \If{$\Allocations_2 \neq \varnothing$}
            \State $\Allocations_1 \leftarrow \Allocations_2$
        \Else
            \For{$\User \in \Trusted \cup \Untrusted$}
                \State $\Allocations_1 \leftarrow \Allocations_1 \cup \connect{\allocation_1, \User, S, C, \Sizes}$
            \EndFor
        \EndIf
    \Else
        \State $\Allocations_1 \leftarrow \newAlloc(\allocation_1, \UserMerge \cup S, \xTalkTriple, C, \Sizes)$
    \EndIf
    \State \Return $\Allocations_1$
\EndProcedure
\State
\Procedure{$\allocTrustedSymb$}{$\allocation_1 = \dataList{\Unallocated, \Trusted, \Untrusted}, \xTalkImpacting, C, \Sizes$}
    \State Set of \typeAllocs $\Allocations \leftarrow \varnothing$
    \For{Set of Qubits $S \subseteq \xTalkImpacting \cap \Unallocated$}
        \For{Set of Qubits $\User \in \Trusted$}
            \If{$\xTalkImpacting \cap (\User \cup S) \neq \varnothing$}
                \State $\Allocations \leftarrow \Allocations \cup \connect{\allocation_1, \User, S, C, \Sizes}$
            \EndIf
        \EndFor
    \EndFor
    \State \Return $\Allocations$
\EndProcedure
\State
\Procedure{$\connectSymb$}{$\allocation_1 = \dataList{\Unallocated, \Trusted, \Untrusted}, \User, S, C, \Sizes$}
    \State Set of \typeAllocs $\Allocations \leftarrow \varnothing$
    \State Integer $\maxLength \leftarrow \remain(\User, \allocation_1, \Sizes) - \card{S \setminus \User}$
    \For{Set of Qubits $\userPath \in \Paths(\User, S, \Unallocated, C, \maxLength)$}
        \State $\Allocations \leftarrow \Allocations \cup \newAlloc(\allocation_1, \User \cup S \cup \userPath, \xTalkTriple, C, \Sizes)$
    \EndFor
    \State \Return $\Allocations$
\EndProcedure
\end{algorithmic}
 \caption{Auxiliary functions used in \Cref{algo:allocation}}
\label{algo:aux3}
\end{figure}

 \fi

\end{document}